\documentclass{article}

\usepackage[utf8]{inputenc}

\usepackage{float}
\usepackage{algorithm}
\usepackage[noend]{algpseudocode}
\usepackage{graphicx}
\usepackage{color}
\usepackage{xy}
\usepackage{url}
\usepackage{afterpage}
\usepackage{fullpage}

\newcommand{\Continue}{\textbf{continue}}
\newcommand{\attr}{\ensuremath{\leftarrow} }
\newcommand{\nullvalue}{\ensuremath{\bot}}
\newcommand{\maxLvl}{\ensuremath{\mathrm{MaxLvl}}}
\newcommand{\maxInt}{\ensuremath{\mathrm{MaxInt}}}

\begin{document}

\title{The Adaptive Priority Queue with Elimination and Combining}
\author{Irina Calciu, Hammurabi Mendes, and Maurice Herlihy\\
Department of Computer Science\\
Brown University\\
115 Waterman St., 4th floor\\
Providence RI -- USA\\
\texttt{\{irina,hmendes,mph\}@cs.brown.edu}
}

\maketitle

\begin{abstract}
Priority queues are fundamental abstract data structures, often used to manage limited resources in parallel programming. Several proposed parallel priority queue implementations are based on skiplists, harnessing the potential for parallelism of the \texttt{add()} operations. In addition, methods such as Flat Combining have been proposed to reduce contention by batching together multiple operations to be executed by a single thread. While this technique can decrease lock-switching overhead and the number of pointer changes required by the \texttt{removeMin()} operations in the priority queue, it can also create a sequential bottleneck and limit parallelism, especially for non-conflicting \texttt{add()} operations.

In this paper, we describe a novel priority queue design, harnessing the scalability of parallel insertions in conjunction with the efficiency of batched removals. Moreover, we present a new elimination algorithm suitable for a priority queue, which further increases concurrency on balanced workloads with similar numbers of \texttt{add()} and \texttt{removeMin()} operations. We implement and evaluate our design using a variety of techniques including locking, atomic operations, hardware transactional memory, as well as employing adaptive heuristics given the workload.
\end{abstract}

\section{Introduction}
\label{Sec-Introduction}

A priority queue is a fundamental abstract data structure that stores a set of keys (or a set of key-value pairs), where keys represent priorities. It usually exports two main operations: \texttt{add()}, to insert a new item in the priority queue, and \texttt{removeMin()}, to remove the first item (the one with the highest priority). Parallel priority queues are often used in discrete event simulations and resource management, such as operating systems schedulers. Therefore, it is important to carefully design these data structures in order to limit contention and improve scalability. Prior work in concurrent priority queues exploited parallelism by using either a heap~\cite{pqhunt} or a skiplist~\cite{Lotan2000} as the underlying data structures. In the skiplist-based implementation of Lotan and Shavit~\cite{Lotan2000} each node has a ``deleted'' flag, and processors contend to mark such ``deleted'' flags concurrently, in the beginning of the list. When a thread logically deletes a node, it tries to remove it from the skiplist using the standard removal algorithm. A lock-free skiplist implementation is presented in \cite{pqsundelltsigas}. 

However, these methods may incur limited scalability at high thread counts due to contention on shared memory accesses. Hendler et al.~\cite{Hendler2010} introduced Flat Combining, a method for batching together multiple operations to be performed by only one thread, thus reducing the contention on the data structure. This idea has also been explored in subsequent work on delegation~\cite{Metreveli2012,CalciuDHHKMM13}, where a dedicated thread called a \emph{server} performs work on behalf of other threads, called \emph{clients}. Unfortunately, the server thread could become a sequential bottleneck. A method of combining delegation with elimination has been proposed to alleviate this problem for a stack data structure~\cite{HotPar13Stack}. Elimination~\cite{Hendler2010a} is a method of matching concurrent inverse operations so that they don't access the shared data structure, thus significantly reducing contention and increasing parallelism for otherwise sequential structures, such as stacks. An elimination algorithm has also been proposed in the context of a queue~\cite{Moir2005}, where the authors introduce the notion of \emph{aging operations} - operations that wait until they become suitable for elimination.

In this paper, we describe, to the best of our knowledge, the first elimination algorithm for a priority queue.
Only \texttt{add()} operations with values smaller than the priority queue minimum value are allowed to eliminate. However, we use the idea of aging operations introduced in the queue algorithm to allow \texttt{add()} values that are \emph{small enough} to participate in the elimination protocol, in the hope that they will soon become eligible for elimination.
We implement the priority queue using a skiplist and we exploit the skiplist's capability for both operations-batching and disjoint-access parallelism. 
\texttt{RemoveMin()} requests can be batched and executed by a server thread using the combining/delegation paradigm. 
 \texttt{Add()} requests with high keys will most likely not become eligible for elimination and need to be inserted in the skiplist, sometimes requiring expensive traversals towards the end of the data structure to do so. Therefore, these operations represent a bottleneck for the server and a missed opportunity for parallelism. To alleviate these issues, we split the underlying skiplist into two parts: a \emph{sequential} part, managed by the server thread and a \emph{parallel} part, where high-valued \texttt{add()} operations can insert their arguments in parallel. Our design reduces contention by performing batched sequential \texttt{removeMin()} and small-value \texttt{add()} operations, while also leveraging parallelism opportunities through elimination and parallel high-value \texttt{add()} operations. We show that our priority queue outperforms prior algorithms in high contention workloads on a SPARC Niagara II machine. Finally, we explore whether the use of hardware transactions could simplify our design and improve throughput. Unfortunately, machines that support hardware transactional memory (HTM) are only available for up to four cores (eight hardware threads), which is not enough to measure scalability of our design in high contention scenarios. Nevertheless, we showed that a transactional version of our algorithm is better than a non-transactional version on a Haswell four-core machine. We believe that these preliminary results will generalize to machines with more threads with support for HTM, once they become available. In summary, our main contributions are:
\begin{itemize}
\item We propose the first elimination algorithm for a priority queue, consisting of (1)~\emph{immediate elimination}, where suitable \texttt{add()} and \texttt{removeMin()} operations exchange arguments; and (2)~\emph{upcoming elimination}, where \texttt{add()} operations with small keys, yet not suitable for elimination, wait some time until either they become suitable or time out.
\item We describe a scalable design for a priority queue based on our elimination algorithm and the delegation/combining paradigm introduced by prior work.
\item We augment our priority queue design with an adaptive component that allows it to perform combining and elimination efficiently, while also allowing \texttt{add()} operations not involved in the elimination to insert in parallel. 
\item We analyze how hardware transactions could be used to simplify and improve our initial design and show performance results on a Haswell machine with transactional memory enabled. 
\end{itemize}

\section{Design}
\label{Sec-Design}

\begin{figure}[htb]
\centering
  \includegraphics[width=0.9\textwidth]{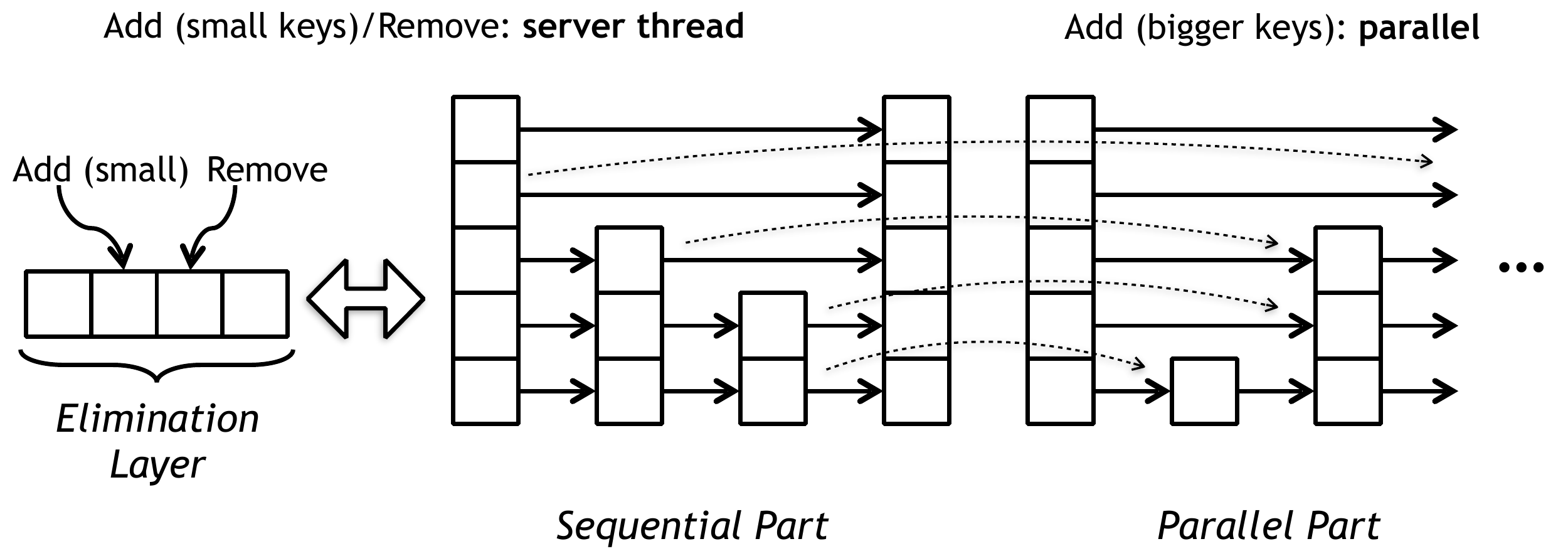}
\caption{Skiplist design. An elimination array is used for \texttt{removeMin()}s and \texttt{add()}s with small keys. A dedicated server thread collects the operations that do not eliminate and executes them on the sequential part of the skiplist. Concurrent threads operate on the parallel part, performing \texttt{add()}s with bigger keys. The dotted lines show pointers that would be established if the single skiplist was not divided in two parts.}
\label{fig:pqe}
\end{figure}

Our priority queue exports two operations: \texttt{add()} and \texttt{removeMin()} and is implemented using an underlying skiplist. The elements of the skiplist are buckets associated with keys. For a bucket $b$, the field $b$.key denotes the associated key.
We split the skiplist in two distinct parts. The sequential part, in the beginning of the skiplist, is likely to serve forthcoming \texttt{removeMin()} operations of the priority queue (\texttt{PQ::removeMin()} for short) as well as \texttt{add($v$)} operations of the priority queue (\texttt{PQ:add()} for short) with $v$ small enough (hence expected to be removed soon). The \emph{parallel part}, which complements the sequential part, is likely to serve \texttt{PQ::add($v$)} operations where $v$ is large enough (hence not expected to be removed soon). Either the sequential or the parallel part may become empty.
Both lists are complete skiplists, with (dummy) head buckets called \texttt{headSeq} and \texttt{headPar}, respectively, with key $- \infty$. Both lists also contain (dummy) \texttt{tail} buckets, with key $+ \infty$. We call the last non-dummy bucket of the sequential part \texttt{lastSeq}, which is the logical divisor between parts. Figure~\ref{fig:pqe} shows the design.

When a thread performs a \texttt{PQ::add($v$)}, either (1) $v > \mathtt{lastSeq.key}$, and the thread inserts the value concurrently in the parallel part of the skiplist, calling the \texttt{SL::addPar()} skiplist operation; or (2) $v \le$ \texttt{lastSeq.key}, and the thread tries to perform elimination with a \texttt{PQ::removeMin()} using the elimination array. 
A \texttt{PQ::add($v$)} with $v$ less than the smallest value in the priority queue can immediately eliminate with a \texttt{PQ::removeMin()}, if one is available. A \texttt{PQ::add($v$)} operation with $v$ bigger than \texttt{minValue} (the current minimal key) but smaller than $\mathtt{lastSeq.key}$ lingers in an \emph{elimination array} for some time, waiting to become eligible for elimination or timeout. A \emph{server thread} executes sequentially all operations that fail to eliminate.

This mechanism describes the first elimination algorithm for a priority queue, well integrated with delegation/combining, presented in more detail in Section~\ref{Sec-Design-EliminationCombining}.
Specifically: (1) The scheme harnesses the parallelism of the priority queue \texttt{add()} operations, letting the ones with keys physically distant and large enough (bigger than \texttt{lastSeq.key}) execute in parallel. (2) At the same time, we batch concurrent priority queue \texttt{add()} with small keys and \texttt{removeMin()} operations that timed out in the elimination array, serving such requests quickly through the server thread -- this latter operation simply consumes elements from the sequential part by navigating through elements in its bottom level, merely decreasing counters and moving pointers in the most common situation. While detaching a sequential part is non-negligible cost-wise, a sequential part has the potential to serve multiple removals.

\subsection{Concurrent Skiplist}
\label{Sec-Design-ConcurrentSkiplist}

Our underlying skiplist is operated by the server thread in the sequential part and by concurrently inserting threads with bigger keys in the parallel part.

\textbf{Sequential part.} 
The server calls the skiplist function \texttt{SL::moveHead()} to extract a new sequential part from the parallel part if some \texttt{PQ::removeMin()} operation was requested and the sequential part was empty. 
Conversely, it calls the skiplist function \texttt{SL::chopHead()} to relink the sequential and the parallel parts, forming a completely parallel skiplist, if no \texttt{PQ::removeMin()} operations are being requested for some time.
In \texttt{SL::moveHead()}, we initially determine the elements to be moved to the sequential part. If no elements are found, the server clears the sequential part, otherwise separating the sequential part from the rest of the list, which becomes the parallel part. The number of elements that \texttt{SL::moveHead()} tries to detach to the sequential part adaptively varies between 8 and 65,536. Our policy is simple: if more than N insertions (e.g. N $=$ 1000) occurred in the sequential part since the last \texttt{SL::moveHead()}, we halve the number of elements moved; otherwise, if less than M insertions (e.g. M $=$ 100) were made, we double this number.
After \texttt{SL::moveHead()} executes, a pointer called \texttt{currSeq} indicates the first bucket in the sequential part, and another called \texttt{lastSeq} indicates the final bucket.
The server uses \texttt{SL::addSeq()} and \texttt{SL::removeSeq()} within the sequential part to remove elements or insert elements with small keys (i.e., belonging to the sequential part) that failed to eliminate.
Buckets are not deleted at this time; they are deleted lazily when the whole sequential part gets consumed. A new sequential part can be created by calling \texttt{SL::moveHead()} again.

\textbf{Parallel part.} The skiplist function \texttt{SL::addPar()} inserts elements into the parallel part, and is called by concurrent threads performing \texttt{PQ::add()}. While these insertions are concurrent, the skiplist still relies on a Single-Writer Multi-Readers lock with writer preference for the following purpose. Multiple \texttt{SL::addPar()} operations acquire the lock for reading (executing concurrently), while \texttt{SL::moveHead()} and \texttt{SL::chopHead()} operations acquire the lock for writing. This way, we avoid that \texttt{SL::addPar()} operates on buckets that are currently being moved to the sequential part by \texttt{SL::moveHead()}, or interferes with \texttt{SL::chopHead()}.
Despite the lock, \texttt{SL::addPar()} is not mutually exclusive with the \emph{head-moving operations} (\texttt{SL::moveHead()} and \texttt{SL::chopHead()}). Only the pointer updates (for new buckets) or the counter increment (for existing buckets) must be done in the parallel part (and not have been moved to the sequential part) after we determine the locations of these changes.
Hence, in the \texttt{SL::addPar()} operation, we first try to get a \emph{clean} \texttt{SL::find()}: a \texttt{SL::find()} followed by acquiring the lock for reading, with no intervening head-moving operations. We can tell whether no head-moving operation took place since our lock operations always increases a timestamp variable, checked in the critical section. After a clean \texttt{SL::find()}, now holding the lock, if a bucket corresponding to the key is found, we insert the element in the bucket (incrementing a counter). Otherwise, a new bucket is created, and inserted level by level using \texttt{CAS()} operations. If a \texttt{CAS()} fails in a certain level, we release the lock and retry a clean \texttt{SL::find()}.

Our algorithm differs from the traditional concurrent skiplist insertion algorithms in two ways: (1) we hold a lock to avoid head-moving operations to take place after a clean \texttt{SL::find()}; and (2) if the new bucket is moved out of the parallel section while we insert the element in the upper levels, we stop \texttt{SL::addPar()}, leaving this element with a capped level. This bucket is likely to be soon consumed by a \texttt{SL::removeSeq()} operation, resulting from a \texttt{PQ::removeMin()} operation.

\subsection{Elimination and Combining}
\label{Sec-Design-EliminationCombining}

Elimination allows matching operations to complete without accessing the shared data structure, thus increasing parallelism and scalability. In a priority queue, any \texttt{SL::removeMin()} operation can be eliminated, but only \texttt{SL::add()} operations with values smaller or equal to the current minimum value can be so. If the priority queue is empty, any \texttt{SL::add()} value can be eliminated. 
We used an elimination array similar to the one in the stack elimination algorithm~\cite{Hendler2010a}. Each slot uses 64 bits to pack together a 32-bit value that represents either an opcode or a value to be inserted in the priority queue and a stamp that is unique for each operation. The opcodes are: EMPTY, REMREQ, TAKEN and INPROG. These are special values that cannot be used in the priority queue. All other values are admissible. In our implementation, each thread has a local count of how many operations it performed. This count is combined with the thread ID to obtain a unique stamp for each operation. Overflow was not an issue in our experiments, but if it becomes a problem a different algorithm for associating unique stamps to each operation could be used. The unique stamp is used to ensure linearizability, as explained in Section~\ref{Sec-Linearizability}. All slots are initially empty, marked with the special value EMPTY, and the stamp value is zero. 
  
A \texttt{PQ::removeMin()} thread loops through the elimination array until it finds a request to eliminate with or it finds an empty slot in the array, as described in Algorithm~\ref{Alg-removeMin}.
If it finds a value in the slot, then it must ensure that the stamp is positive, otherwise the value was posted as a response to another thread. The value it finds must be smaller than the current priority queue minimum value. Then, the \texttt{PQ::removeMin()} thread can \texttt{CAS} the slot, which contains both the value and the stamp, and replace it with an indicator that the value was taken (TAKEN, with stamp zero). The thread returns the value found. If instead, the \texttt{PQ::remove()} thread finds an empty slot, it posts a \emph{remove request} (REMREQ), with a unique stamp generated as above. The thread waits until the slot is changed by another thread, having a value with stamp zero. The \texttt{PQ::removeMin()} thread can then return that value. 

\begin{algorithm}[htb]
\caption{PQ::removeMin()}
\label{Alg-removeMin}
\begin{algorithmic}[1]
\While{$\mathbf{true}$}
	\State pos \attr $(id + 1) \% $ ELIM\_SIZE; (value, stamp) \attr elim[pos]
	\If {IsValue(value) \textbf{and} (stamp $> 0$) \textbf{and} (value $\le$ skiplist.minValue))}
	    \If{CAS(elim[pos], (value, stamp), (TAKEN, 0))}
		\State \Return value
	    \EndIf
	\EndIf
	\If {value = EMPTY}
	    \If{CAS(elim[pos], (value, stamp), (REMREQ, uniqueStamp()))}
		\Repeat
		  \State (value, stamp) \attr elim[pos]
		\Until{value $\ne$ REMREQ \textbf{and} value $\ne$ INPROG}
		\State elim[pos] \attr (EMPTY, 0); \Return value
	    \EndIf    
	\EndIf
	\State inc(pos)
\EndWhile
\end{algorithmic}
\end{algorithm}

A \texttt{PQ::add()} thread initially tries to use \texttt{SL::addPar()} to add its value in parallel. A failed attempt to add in parallel indicates that the value should try to eliminate or should be inserted in the sequential part. The \texttt{PQ::add()} thread tries to eliminate by checking through the elimination array for REMREQ indicators. If it finds a remove request, and its value is smaller than the priority queue \texttt{minValue}, it can \texttt{CAS} its value with stamp zero, effectively handing it to another thread. If multiple such attempts fail, the thread changes its behavior: it still tries to perform elimination as above, but as soon as an empty slot is found, it uses a \texttt{CAS} to insert its own value and the current stamp in the slot, waiting for another thread to match the operation (and change the opcode to TAKEN) returning the corresponding value.

\begin{figure}
  \centering
	  \includegraphics[width=0.75\textwidth]{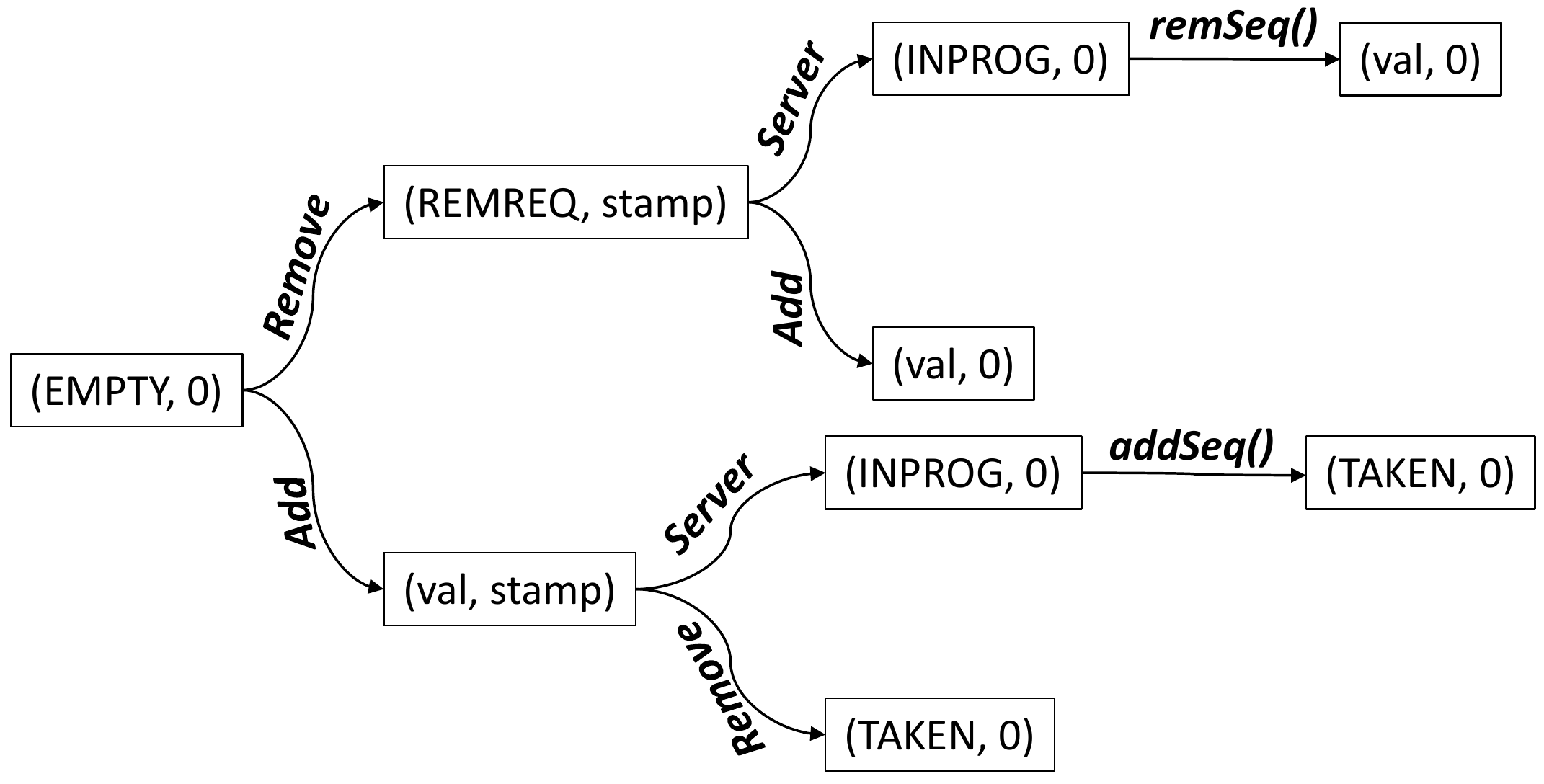}
\caption{Transitions of a slot in the elimination array.}
\label{fig:transitions}
\end{figure}

The \texttt{PQ::add()} and \texttt{PQ::removeMin()} threads that post a request in an empty slot of the elimination array wait for a matching thread to perform elimination. However, elimination could fail because no matching thread shows up or because the \texttt{PQ::add()} value is never smaller than the priority queue \texttt{minValue}. To ensure that all threads make progress, we use a dedicated \emph{server thread} that collects add and remove requests that fail to eliminate. 
The server thread executes the operations sequentially on the skiplist, calling \texttt{SL::addSeq()} and \texttt{SL::removeSeq()} operations. To ensure linearizability, the server marks a slot that contains an operation it is about to execute as \emph{in progress} (INPROG). Subsequently, it executes the sequential skiplist operation and writes back the response in the elimination slot for the other thread to find it. A state machine showing the possible transitions of a slot in the elimination array is shown in Figure~\ref{fig:transitions}, and the algorithm is described in Algorithm~\ref{Alg-Execute}.

\begin{algorithm}[htb]
\caption{Server::execute()}
\label{Alg-Execute}
\begin{algorithmic}[1]
\While{$\mathbf{true}$}
  \For{$i$: 1 $\rightarrow$ ELIM\_SIZE}
	\State (value, stamp) \attr elim[i]
	\If{value = REMREQ}
	    \If{CAS(elim[i], (value, stamp), (INPROG, 0))}
		\State min \attr skiplist.removeSeq(); elim[i] \attr (min, 0)
	    \EndIf
	\EndIf
	\If {IsValue(value) \textbf{and} (stamp $> 0$)}
	    \If{CAS(elim[i], (value, stamp), (INPROG, 0))}
		\State skiplist.addSeq(value); elim[i] \attr (TAKEN, 0)
	    \EndIf    
	\EndIf
  \EndFor
\EndWhile
\end{algorithmic}
\end{algorithm}

\section{Linearizability}
\label{Sec-Linearizability}
 
Our design provides a linearizable priority queue algorithm. Some operations have multiple possible linearization points by design, requiring careful analysis and implementation.

\textbf{Skiplist.} A successful \texttt{SL::addPar($v$)} (respectively, \texttt{SL::addSeq($v$)}) usually linearizes when it inserts the element in the bottom level of the skip list with a \texttt{CAS} (respectively, with a store), or when the bucket for key $v$ has its counter incremented with a \texttt{CAS} (respectively, with a store). However, a thread inserting a minimal bucket, whenever $v < \mathtt{minValue}$, is required to update \texttt{minValue}. When the sequential part is not empty, only the server can update \texttt{minValue} (without synchronization). When the sequential part is empty, a parallel add with minimal value needs to update \texttt{minValue}.
The adding thread loops until a \texttt{CAS} decreasing \texttt{minValue} succeeds or another thread inserts a bucket with key smaller than $v$. Note that no head-moving operation is taking place (the \texttt{SL::addPar()} threads hold the \texttt{lock}). Threads that succeed changing \texttt{minValue} linearize their operation at the point of the successful \texttt{CAS}. 

The head-moving operations \texttt{SL::moveHead()} and \texttt{SL::chopHead()} execute while holding the \texttt{lock} for writing, which effectively linearizes the operation at the \texttt{lock.release()} instant because: (1) no \texttt{SL::addPar()} is running; (2) no \texttt{SL::addSeq()} or \texttt{SL::removeSeq()} are running, as the server thread is the single thread performing those operations. Head-moving operations do not change \texttt{minValue}, in fact they preclude any changes to it. During these operations, however, threads may still perform elimination, which we discuss next.

\begin{figure}
  \centering
  \includegraphics[width=0.9\textwidth]{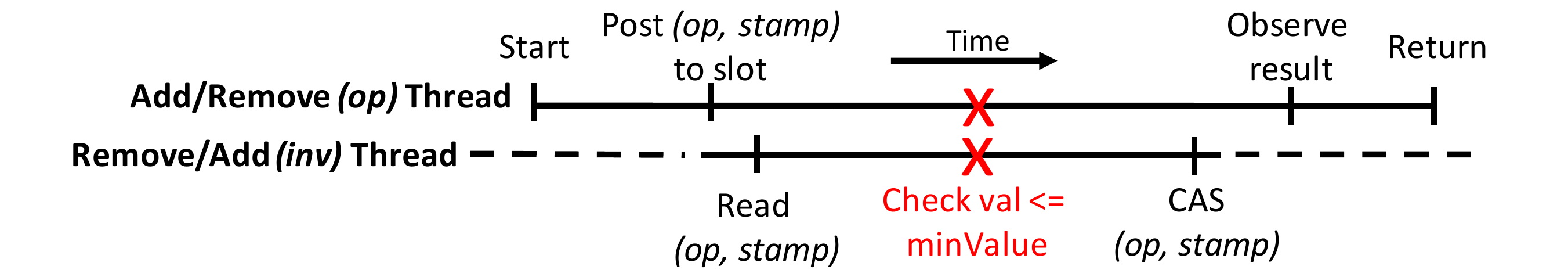}
\caption{Concurrent execution of an \emph{op} thread posting its request to an empty slot, and an \emph{inv} thread, executing a matching operation. The operation by the \emph{inv} thread could begin anytime before the \emph{Read} and finish any time after the \emph{CAS}. The linearization point is marked with a red X.}
\label{fig:correctness_elim}
\end{figure}

\textbf{Elimination.} A unique stamp is used in each request posted in the array entries to avoid the ``ABA'' problem. 
Each elimination slot is a 64-bit value that contains 32 bits for the posted value (for \texttt{PQ::add()}) or a special opcode (for \texttt{PQ::removeMin()}) and 32 bits for the unique stamp. In our implementation, the unique stamp is obtained by combining the thread id with the number of operations performed by each thread. 
Each thread, either adding or removing, that finds the inverse operation in the elimination array must verify that the exchanged value is smaller than \texttt{minValue}. If so, the thread can \texttt{CAS} the elimination slot, exchanging arguments with the waiting thread. It is possible that the priority queue minimum value is changed by a concurrent \texttt{PQ::add()}. In that case, the linearization point for both threads engaged in  elimination is at the point where the value was observed to be smaller than the priority queue minimum. See Fig.~\ref{fig:correctness_elim}.

The thread performing the \texttt{CAS} first reads the stamp of the thread that posted the request in the array and verifies that it is allowed to eliminate. Only then it performs a \texttt{CAS} on both the value and the stamp, guaranteeing that the thread waiting did not change in the meantime. Because both threads were running at the time of the verification, they can be linearized at that point. Without the unique stamp, the eliminating thread could perform a \texttt{CAS} on an identical request (i.e., identical operation and value) posted in the array by a different thread. The \texttt{CAS} would incorrectly succeed, but the operations would not be linearizable because the new thread was not executing while the suitable minimum was observed.

\begin{figure}[htb]
  \centering
  \includegraphics[width=0.9\textwidth]{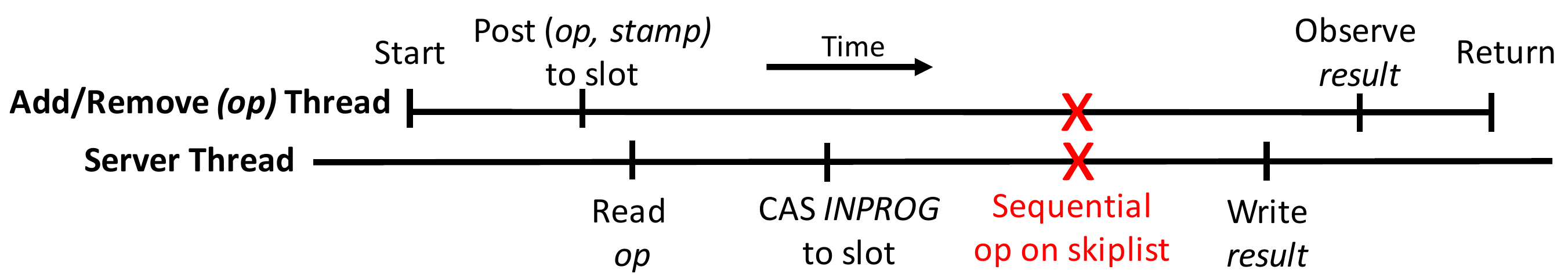}
\caption{Concurrent execution of a client thread and the server thread. The client posts its operation \emph{op} to an empty slot and waits for the server to collect the operation and execute it sequentially on the skiplist. The linearization point occurs in the sequential operation and is marked with a red X.}
\label{fig:correctness_server}
\end{figure}

The linearizability of the combining operation results from the linearizability of the skiplist. The threads post their operation in the elimination array and wait for the server to process it. The server first marks the operation as \emph{in progress} by \texttt{CAS}ing INPROG into the slot. Then it performs the sequential operation on the skiplist and writes the results back in the slot, releasing the waiting thread. The waiting thread observes the new value and returns it. The linearization point of the operation happens during the sequential operation on the skiplist, as discussed above. See Fig.~\ref{fig:correctness_server}.

\section{Evaluation}
\label{Sec-Evaluation}

In this section, we discuss results on a Sun SPARC T5240, which contains two UltraSPARC T2 Plus chips with 8 cores each, running at 1.165 GHz. Each core has 8 hardware strands, for a total of 64 hardware threads per chip. A core has a 8KB L1 data cache and shares an 4MB L2 data cache with the other cores on a chip. Each experiment was performed five times and we report the median. Variance was very low for all experiments. Each test was run for ten seconds to measure throughput. We used the same benchmark as flat combining~\cite{Hendler2010}. A thread randomly flips a coin with  probability $p$ to be an \texttt{PQ::add()} and $1-p$ to be a \texttt{PQ::removeMin()}. We started a run after inserting 2000 elements in the priority queue for stable state results. 

Our priority queue algorithm (\emph{pqe}) uses combining and elimination, and leverages the parallelism of \texttt{PQ::add()}. We performed experiments to compare against previous priority queues using combining methods, such as flat combining skiplist (\emph{fcskiplist}) and flat combining pairing heap (\emph{fcpairheap}). We also compared against previous priority queues using skiplists with parallel operations, such as a lock free skiplist (\emph{lfskiplist}) and a lazy skiplist (\emph{lazyskiplist}). The flat combining methods are very fast at performing \texttt{PQ::removeMin()} operations, which then get combined and executed together. However, performing the \texttt{PQ::add()} operations sequentially is a bottleneck for these methods. Conversely, the \emph{lfskiplist} and \emph{lazyskiplist} algorithms are very fast at performing the parallel adds, but get significantly slowed down by having \texttt{PQ::removeMin()} operations in the mix, due to the synchronization overhead involved. Our \emph{pqe} design tries to address these limitations through our \emph{dual} (sequential and parallel parts), \emph{adaptive} implementation that can be beneficial in the different scenarios. 

\begin{figure}[htb]
\centering
\begin{minipage}[b]{.495\textwidth}
	\centering
  \includegraphics[width=\linewidth]{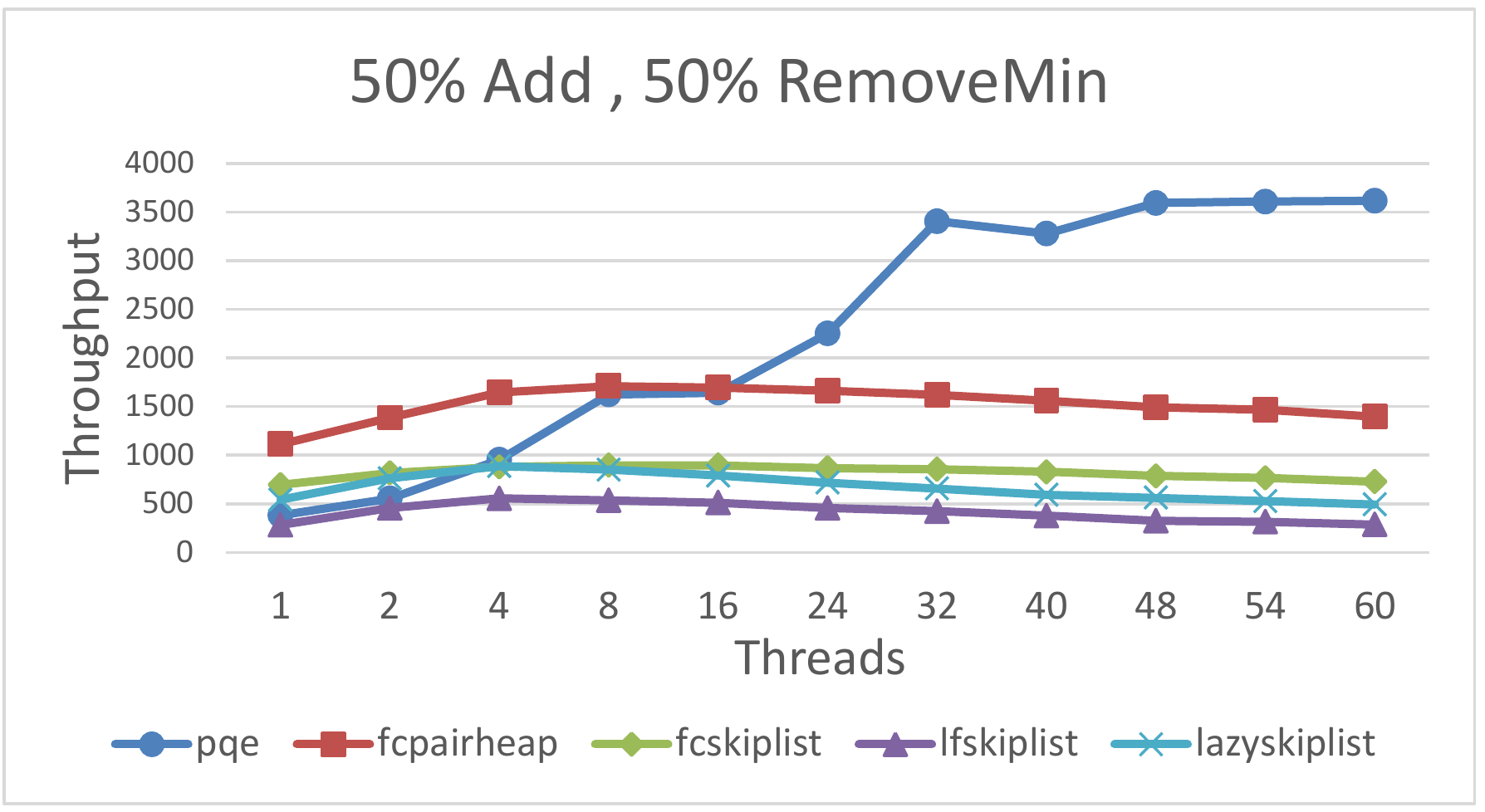}
\caption{Priority queue performance with 50\% \texttt{add()}s, 50\% \texttt{removeMin()}s.}
\label{fig:sparc_50}
\end{minipage}%
\hfill%
\begin{minipage}[b]{.495\textwidth}
	\centering
  \includegraphics[width=\linewidth]{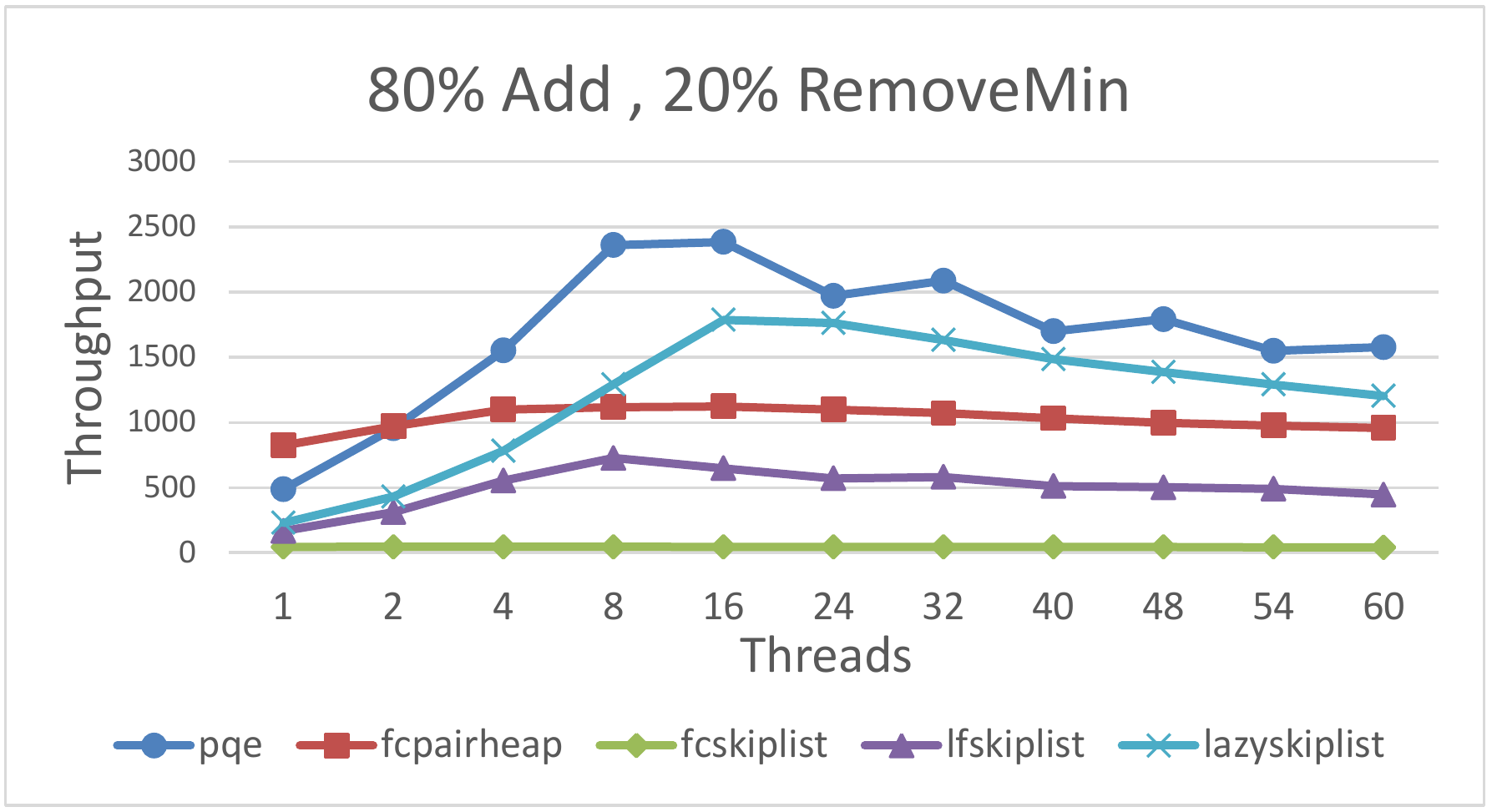}
\caption{Priority queue performance with 80\% \texttt{add()}s, 20\% \texttt{removeMin()}s.}
\label{fig:sparc_80}
\end{minipage}%
\end{figure}

\begin{figure}[htb]
\centering
\begin{minipage}[b]{.495\textwidth}
	\centering
  \includegraphics[width=\linewidth]{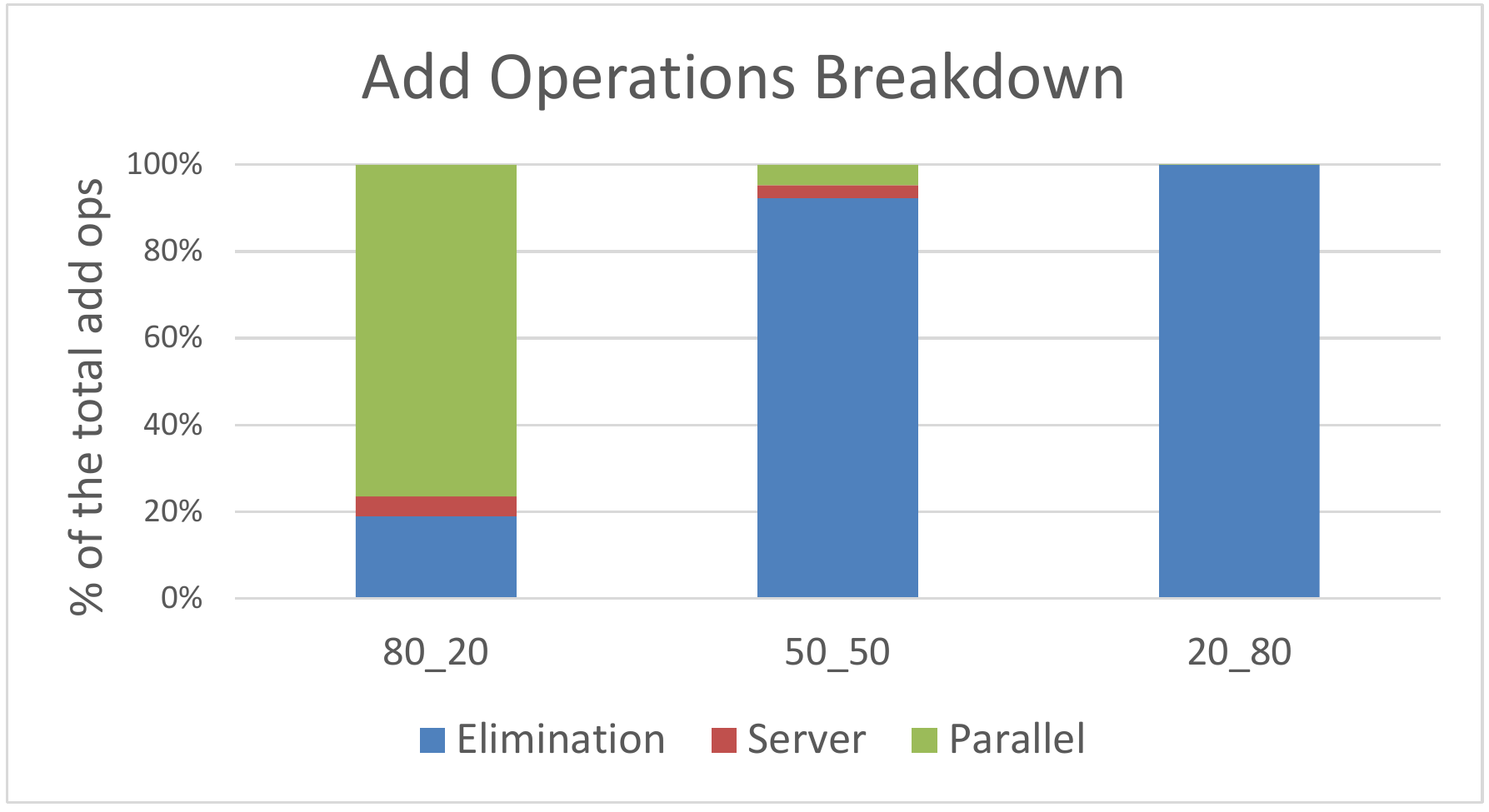}
\caption{\texttt{add()} work breakdown.}
\label{fig:sparc_add}
\end{minipage}%
\hfill%
\begin{minipage}[b]{.495\textwidth}
	\centering
  \includegraphics[width=\linewidth]{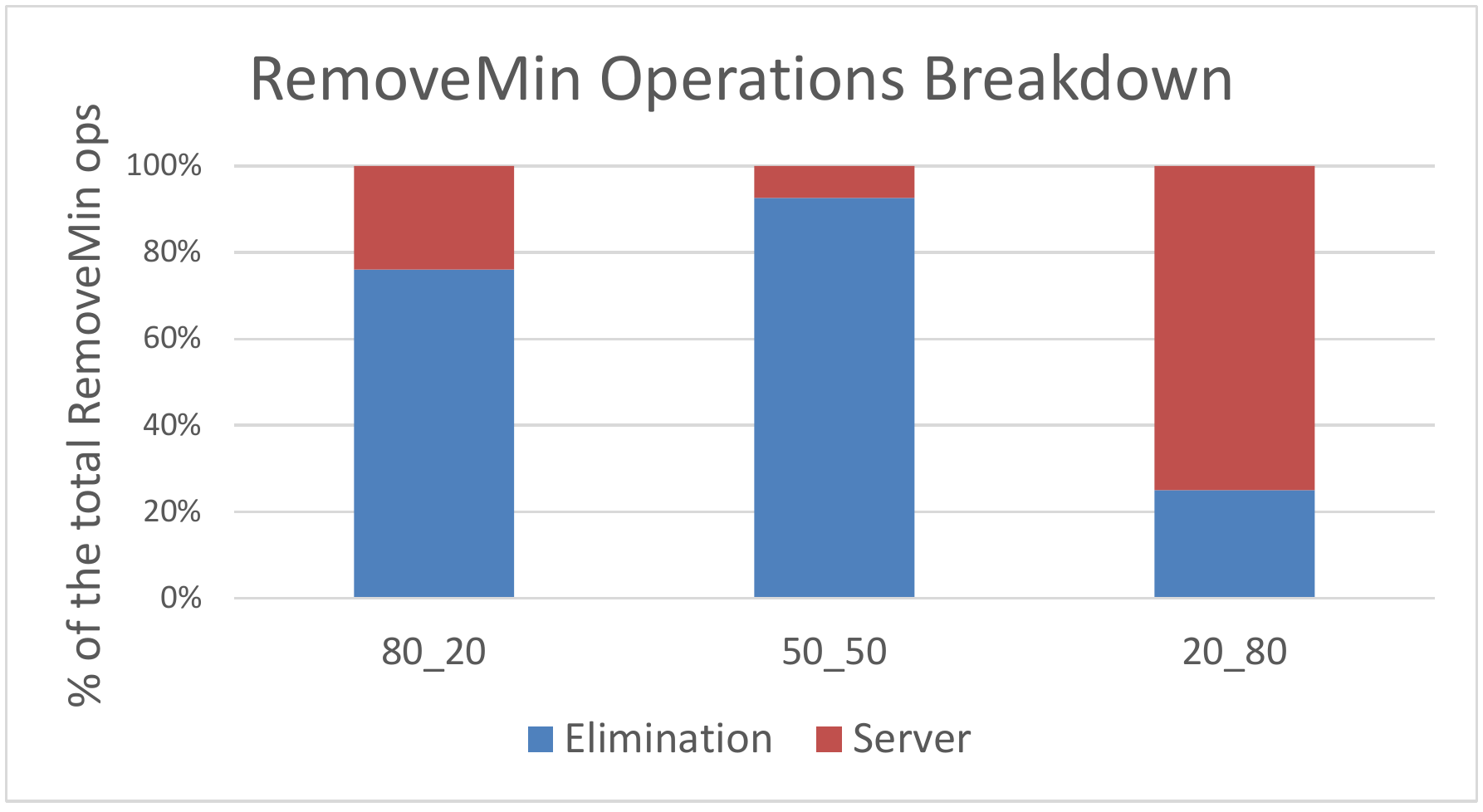}
\caption{\texttt{removeMin()} work breakdown.}
\label{fig:sparc_rem}
\end{minipage}%
\end{figure}

We considered different percentages of \texttt{PQ::add()} and \texttt{PQ::removeMin()} in our tests. When the operations are roughly the same number, \emph{pqe} can fully take advantage of both elimination and parallel adds, so it has peak performance. Figure~\ref{fig:sparc_50} shows how for $50\%$ \texttt{PQ::add()} and $50\%$ \texttt{PQ::removeMin()}, \emph{pqe} is much more scalable and can be up to 2.3 times faster than all other methods. When there are more \texttt{PQ::add()} than \texttt{PQ::removeMin()}, as in Figure~\ref{fig:sparc_80} with $80\%$ \texttt{PQ::add()} and $20\%$ \texttt{PQ::removeMin()}, \emph{pqe} behavior approaches the other methods, but it is still $70\%$ faster than all other methods at high thread counts. In this specific case there is only little potential for elimination, but having parallel insertion operations makes our algorithm outperform the flat combining methods. The \emph{lazyskiplist} algorithm also performs better than other methods, as it also takes advantage of parallel insertions. However, \emph{pqe} uses the limited elimination and the combining methods to reduce contention, making it faster than the \emph{lazyskiplist}. For more \texttt{PQ::removeMin()} operations than \texttt{PQ::add()} operations, the \emph{pqe}'s potential for elimination and parallel adds are both limited, thus other methods can be faster. \emph{Pqe} is designed for high contention scenarios, in which elimination and combining thrive. Therefore, it can incur a penalty at lower thread counts, where there is not enough contention to justify the overhead of the indirection caused by the elimination array and the server thread. 

To better understand when each of the optimizations used is more beneficial, we analyzed the breakdown of the \texttt{PQ::add()} and \texttt{PQ::removeMin()} operations for different \texttt{PQ::add()} percentages. When we have $80\%$ \texttt{PQ::add()}, most of them are likely to be inserted in parallel ($75\%$), with a smaller percentage being able to eliminate and an even smaller percentage being executed by the server, as shown in Fig.~\ref{fig:sparc_add}. In the same scenario, $75\%$ of \texttt{removeMin()} operations eliminate, while the rest gets execute by the server, as seen in Fig.~\ref{fig:sparc_rem}. For balanced workloads ($50\%-50\%$), most operations eliminate and a few \texttt{PQ::add()} operations are inserted in parallel. When the workload is dominated by \texttt{PQ::removeMin()}, most \texttt{PQ::add()} eliminate, but most \texttt{PQ::removeMin()} are still left to be executed by the server thread, thus introducing a sequential bottleneck. Eventually the priority queue would become empty, not being able to satisfy \texttt{PQ::removeMin()} requests with an actual value anymore. In this case, any \texttt{add()} operation can eliminate, allowing full parallelism. We do not present results for this case because it is an unlikely scenario that unrealistically favors elimination. 

\subsection{Evaluating the Overhead of \texttt{PQ::moveHead()} and \texttt{PQ::chopHead()}}
\label{Sec-Eval-Move}

Maintaining separate skiplists for the sequential and the parallel part of the priority queue is beneficial for the overall throughput, but adds some overhead, which we quantify in this section. 
The number of elements that become part of the sequential skiplist changes dynamically based on the observed mix of operations. This adaptive behavior helps reduce the number of \texttt{moveHead()} and \texttt{chopHead()} operations required.  
Table~\ref{fig:sparc_headmove} shows the percentage of the number of head-moving operations out of the total number of \texttt{PQ::removeMin()} operations for different mixes of \texttt{PQ::add()} and \texttt{PQ::removeMin()} operations. The head-moving operations are rarely called due to the priority queue's adaptive behavior. 

\begin{table}[htb]
  \centering
	\includegraphics[width=0.65\textwidth]{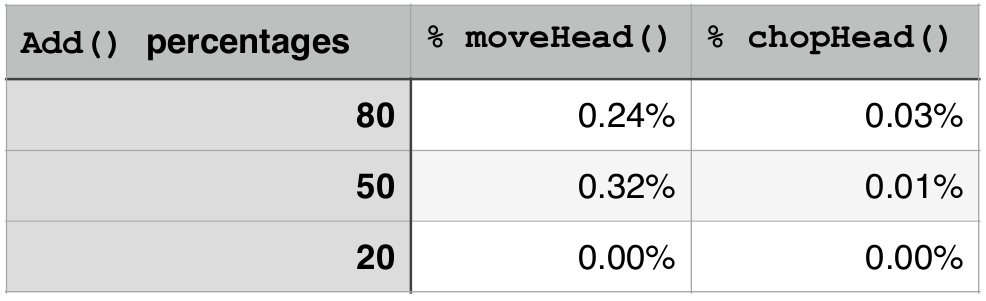}
	\caption{The number of head-moving operations as a percentage of the total number of \texttt{PQ::removeMin()} operations, considering different \texttt{add()} and \texttt{removeMin()} mixes.}
\label{fig:sparc_headmove}
\end{table}

\section{Hardware Transactions}
\label{Sec-HardwareTransactions}


Transactional memory~\cite{Herlihy:1993:TMA:173682.165164} is an optimistic mechanism to synchronize threads accessing shared data. Threads are allowed to execute critical sections speculatively in parallel, but, if there is a data conflict, one of them has to roll back and retry its critical section. Recently, IBM and Intel added HTM instructions to their processors~\cite{wang:2012:pact,haswell:2012:rtm}. 
In our priority queue implementation, we used Intel's Transactional Synchronization Extensions (TSX)~\cite{haswell:2012:rtm} to simplify the implementation and reduce the overhead caused by the synchronization necessary to manage a sequential and a parallel skiplist. 
We evaluate our results on an Intel Haswell four core processor, Core i7-4770, with hardware transactions enabled (restricted transactional memory - RTM), running at 3.4GHz. There are 8GB of RAM shared across the machine and each core has a 32KB L1 cache. Hyperthreading was enabled on our machine so we collected results using all 8 hardware threads. Hyperthreading causes resource sharing between the hyperthreads, including L1 cache sharing, when running with more than 4 threads, thus it can negatively impact results, especially for hardware transactions. We did not notice a hyperthreading effect in our experiments. We used the GCC 4.8 compiler with support for RTM and optimizations enabled (-O3).  

\subsection{Skiplist}
\label{Sec-Transactions-SkipList}

The Single-Writer-Multi-Readers lock used to synchronize the sequential and the parallel skiplists complicates the priority queue design and adds overhead. In this section, we explore an alternative design using hardware transactions. 
However, the naive approach of making all operations transactional causes too many aborts. Instead, the server increments a timestamp whenever a head-moving operation - \texttt{SL::moveHead()} or \texttt{SL::chopHead()} - starts or finishes. A \texttt{SL::addPar()} operation first reads the timestamp and executes a nontransactional \texttt{SL::find()} and then starts a transaction for the actual insertion, adding the server's timestamp to its read set and aborting if it is different from the initially recorded value. Moreover, if the timestamp changes after starting the transaction, indicating a head-moving operation, the transaction will be aborted due to the timestamp conflict.  
If the timestamp is valid, \texttt{SL::find()} must have recorded the predecessors and successors of the new bucket at each level $i$ in \texttt{preds[i]} and \texttt{succs[i]}, respectively. If a bucket already exists, the counter is incremented inside the transaction and the operation completes. If the bucket doesn't exist, the operation proceeds to check if \texttt{preds[i]} points to \texttt{succs[i]} for all levels $0 \le i \le \maxLvl$. If so, the pointers have not changed before starting the transaction and the new bucket can be correctly inserted between \texttt{preds[i]} and \texttt{succs[i]}. Otherwise, we commit the (innocuous) transaction, yet restart the operation.

\begin{figure}[htb]
\centering
\begin{minipage}{.495\textwidth}
	\centering
  \includegraphics[width=\linewidth]{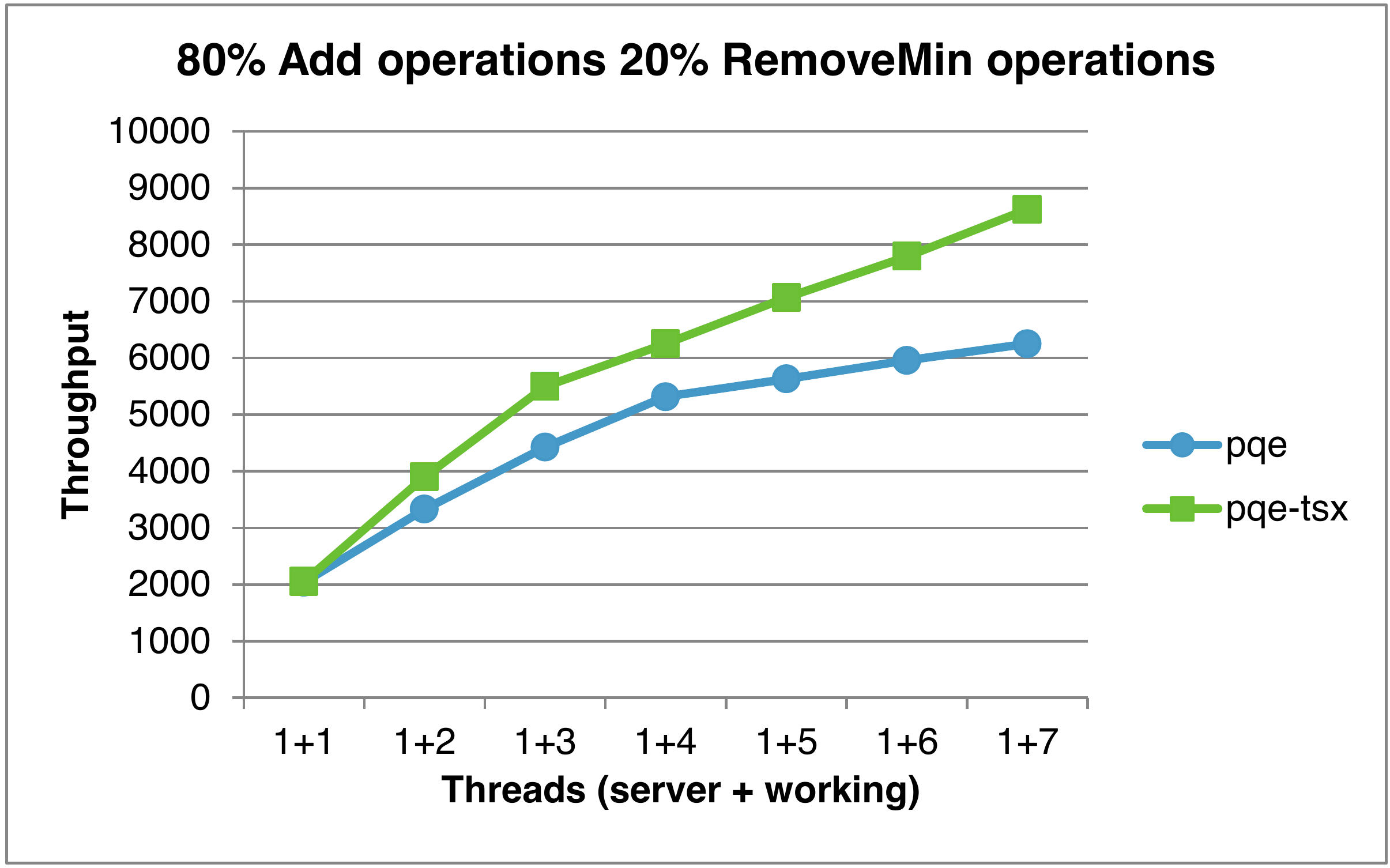}
\caption{Priority queue performance when we use a transaction-based dual skiplist; 80\% \texttt{add()}s, 20\% \texttt{removeMin()}s.}
\label{fig:tsx1}
\end{minipage}%
\hfill%
\begin{minipage}{.495\textwidth}
	\centering
  \includegraphics[width=\linewidth]{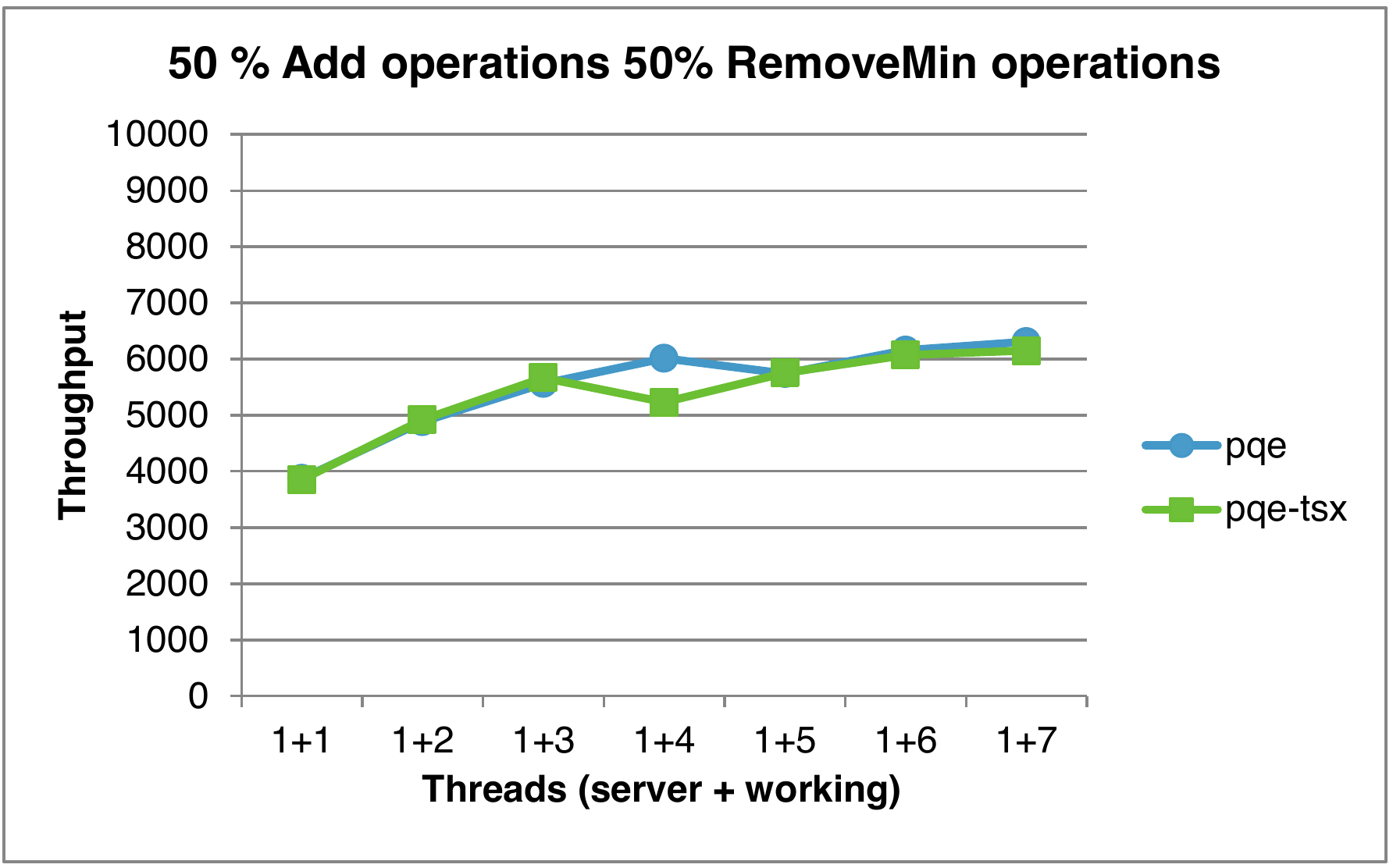}
\caption{Priority queue performance when we use a transaction-based dual skiplist; 50\% \texttt{add()}s, 50\% \texttt{removeMin()}s.}
\label{fig:tsx2}
\end{minipage}
\end{figure}

Figures~\ref{fig:tsx1} and~\ref{fig:tsx2} compare the performance of the lock-based implementation and the implementation based on hardware transactions for two different percentages of \texttt{PQ::add()}s and \texttt{PQ::removeMin()}s. When fewer \texttt{PQ::removeMin()} operations are present, the timestamp changes less frequently and the \texttt{PQ::add()} transactions are aborted fewer times, which increases performance  in the 80\%-20\% insertion-removal mix. In the $50\%$-$50\%$ mix, we obtain results comparable to the \emph{pqe} algorithm using the lock-based approach, albeit with a much simpler implementation.

\subsection{Evaluating the Overhead of Aborted Transactions}
\label{Sec-Eval-Aborted}

The impact of aborted transactions is reported in Tables~\ref{tbl:tsx-stat1} and~\ref{tbl:tsx-stat2}. As the number of threads increases, the number of transactions per successful operation 
also increases, as does the percentage of operations that need more than 10 retries to succeed. 
Note that the innocuous transactions that find inconsistent pointers, changed between the \texttt{SL::find()} and the start of the transaction are not included in the measurement. 
After 10 retries, threads give up on retrying the transactional path and the server executes the operations on their behalf, either in the sequential part, using sequential operations, or in the parallel part, using \texttt{CAS()} for the pointer changes,  but without holding the readers lock. The server does not need to acquire the readers lock because no other thread will try to acquire the writer lock.

\begin{table}[htb]
\centering
\begin{minipage}{.49\textwidth}
	\centering
  \includegraphics[width=\linewidth]{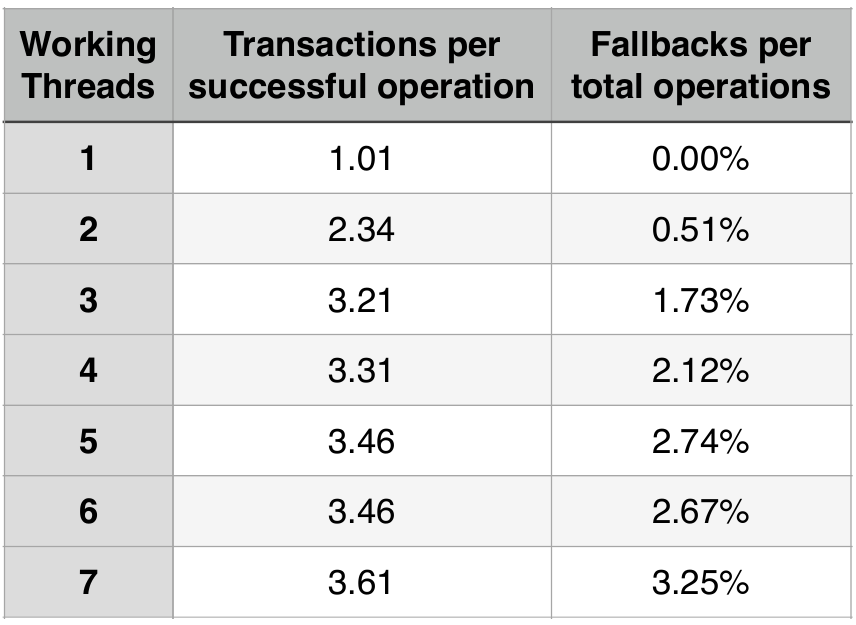}
\caption{Transaction stats for varying \# of threads, with 50\% \texttt{PQ::add()}s and 50\% \texttt{PQ::removeMin()}s}
\label{tbl:tsx-stat1}
\end{minipage}%
\hfill%
\begin{minipage}{.49\textwidth}
	\centering
  \includegraphics[width=\linewidth]{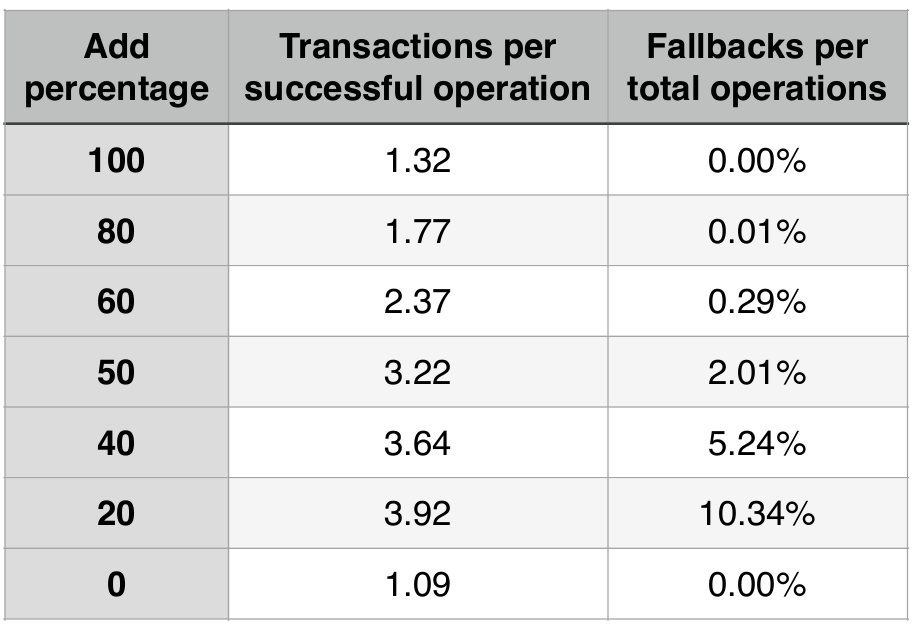}
\caption{Transaction stats for varying mixes, with 1 server thread and 3 working threads.}
\label{tbl:tsx-stat2}
\end{minipage}
\caption{Statistics on the overhead of aborted transactions.}
\end{table}

The number of transactions per successful operation is at most $3.92$, but $3.22$ in the $50\%-50\%$ case. The percentage of operations that get executed by the server (after aborting 10 times) is at most $10\%$ of the total number of operations, but between $1.73\%$ and $2.01\%$ for the $50\%-50\%$ case.

\section{Conclusion}
\label{Sec-Conclusion}

In this paper, we describe a technique to implement a scalable, linearizable priority queue based on a skiplist, divided into a sequential and a parallel part. Our scheme simultaneously enables parallel \texttt{PQ::add()} operations as well as sequential batched \texttt{PQ::removeMin()} operations. The sequential part is beneficial for batched removals, which are performed by a special \emph{server thread}. While detaching the sequential part from the parallel part is non-negligible cost-wise, the sequential part has the potential to serve multiple subsequent removals at a small constant cost. 
The parallel part is beneficial for concurrent insertions of elements with bigger keys (smaller priority), not likely to be removed soon. 
In other words, we integrate the flat combining/delegation paradigm introduced in prior work with disjoint-access parallelism.

In addition, we present a novel priority queue elimination algorithm, where \texttt{PQ::add()} operations with keys smaller than the priority queue minimum can eliminate with \texttt{PQ::removeMin()} operations. We permit \texttt{PQ::add()} operations, with keys small enough, to linger in the elimination array, waiting to become eligible for elimination. If the elimination is not possible, the operation is delegated to the server thread. Batched removals (combining) by the server thread is well-integrated with both: (1) parallelism of \texttt{add()} operations with bigger keys; and (2) the elimination algorithm, that possibly delegates failed elimination attempts (of elements with smaller keys) to the server thread in a natural manner. Our priority queue integrates delegation, combining, and elimination, while still leveraging the parallelism potential of insertions.

\bibliography{pqe_disc_14}
\bibliographystyle{plain}

\newpage
\appendix

\section{Algorithms for the Concurrent Skiplist}
\label{App-Algorithms-ConcurrentSkiplist}

In this section, we present the algorithms for the concurrent skiplist described in Sec.~\ref{Sec-Design-ConcurrentSkiplist}. The skiplist contains a Single-Writer-Multi-Readers lock with writer preference, called simply \texttt{lock}. In terms of notation, \texttt{lock.acquireR()} acquires the lock for reads, and \texttt{lock.acquireW()} acquires the lock for writes. The \texttt{SL::removeSeq()} skiplist procedure is described in Alg.~\ref{Alg-RemoveSeq}.

\begin{algorithm}[!htb]
\caption{SL::removeSeq()}
\label{Alg-RemoveSeq}
\begin{algorithmic}[1]
\If{minValue $=$ \maxInt}
	\State \Return \maxInt
\EndIf
\If{currSeq $=$ \nullvalue}
	\State moveHead()
\EndIf
\State key \attr currSeq.key
\State currSeq.counter \attr currSeq.counter - 1
\If{currSeq.counter $= 0$}
	\While{currSeq $\ne$ lastSeq}
		\State currSeq \attr currSeq.next[0]
		\If{currSeq.counter $> 0$}
			\State minValue = currSeq.key
			\State \Return key
		\EndIf
	\EndWhile
	\State moveHead()
\EndIf
\State \Return key
\end{algorithmic}
\end{algorithm}

The variable \texttt{lock.timestmap} contains the timestamp associated with the lock (and hence with the head-moving operations). Algorithm~\ref{Alg-CleanFind} returns a pair of elements $(b,r)$: $b$ is a bucket found using the skiplist \texttt{SL::find()} operation, and $r$ is a boolean defined as follows. If a head-moving operation happened anywhere between Lines~\ref{algFLT:checkstamp1} and~\ref{algFLT:checkstamp2}, the timestamp moved and $r$ will be false.

\begin{algorithm}[!htb]
\caption{cleanFind($v$, preds, succs)}
\label{Alg-CleanFind}
\begin{algorithmic}[1]
\State $t$ \attr lock.timestamp \label{algFLT:checkstamp1}
\State $b$ \attr find(headPar, $v$, preds, succs)
\State lock.acquireR()
\If{$t$ $<$ lock.timestamp} \label{algFLT:checkstamp2}
	\State lock.release()
	\State \Return $(\nullvalue, \mathbf{false})$
\EndIf
\State \Return $(b, \mathbf{true})$
\end{algorithmic}
\end{algorithm}

The \texttt{SL::addPar()} skiplist procedure is described in Alg.~\ref{Alg-AddPar}. It uses the clean find protocol above. It performs a clean find, followed by mutable operations (either increasing a counter or inserting a bucket), executed with \texttt{lock} acquired for reading.

\begin{algorithm}[!htb]
\caption{SL::addPar($v$)}
\label{Alg-AddPar}
\begin{algorithmic}[1]
\If{$v$ $\le$ lastSeq.key}
	\State \Return \textbf{false}
\EndIf
\State $(b,r)$ \attr cleanFind($v$, preds, succs) \label{algAddPar:find}
\If{$r$ = \textbf{false}}
	\State \textbf{restart} at line~\ref{algAddPar:find}
\EndIf
\If{$b \ne \nullvalue$} \label{algAddPar:incblock}
	\State Atomically increment $b$.counter
	\State lock.release()
	\State \Return \textbf{true}
\EndIf
\State $b$ \attr newNode($v$)
\For{$i$: 1 $\rightarrow$ $b$.topLevel}
	\State $b$.next[i] \attr succs[i]
\EndFor
\If{\textbf{not} CAS(preds[0].next[0]: succs[0] $\rightarrow$ $b$)}
	\State lock.release()
	\State \textbf{restart} at line~\ref{algAddPar:find}
\EndIf
\Repeat
	\State $m$ \attr minValue
\Until{$m \le v$ \textbf{or} CAS(minValue: $m \rightarrow v$)}
\For{$i$: 1 $\rightarrow$ $b$.topLevel}
	\State $b$.next[i] \attr succs[i]
	\If{CAS(preds[i].next[i]: succs[i] $\rightarrow$ $b$)}
		\State \Continue
	\EndIf
	\State lock.release()
	\Repeat
		\State $(b,r)$ \attr cleanFind($v$, preds, succs)
	\Until{$r = \mathbf{true}$}
	\If{$b = \nullvalue$}
		\State lock.release()
		\State \Return \textbf{true}
	\EndIf
\EndFor
\State \Return \textbf{true}
\end{algorithmic}
\end{algorithm}

The \texttt{SL::moveHead()} skiplist procedure is described in Alg.~\ref{Alg-MoveHead}. Line~\ref{algMH:copyHead} creates the sequential part starting from where the parallel part used to be, and the operations starting at Line~\ref{algMH:unlink} separate the skiplist in two parts. Note how \texttt{SL::find()} is used to locate the pointers that will change in order to separate the skiplist.

\begin{algorithm}[!htb]
\caption{SL::moveHead()}
\label{Alg-MoveHead}
\begin{algorithmic}[1]
\State $n$ is determined dynamically (see text)
\State lock.acquireW()
\State currSeq \attr \nullvalue

\State pred \attr headPar
\State curr \attr headPar.next[0]
\State $i = 0$
\While{$i < n$ \textbf{and} curr $\ne$ tail} \label{algMH:consume}
	\State i \attr i + curr.counter
	\If{currSeq = \nullvalue}
		\State currSeq \attr curr; minValue \attr curr.key
	\EndIf
	\State pred \attr curr; curr \attr curr.next[0]
\EndWhile

\If{$i = 0$}
	\For{$i: \maxLvl \rightarrow 0$}
		\State headPar[i], headSeq[i] \attr tail
	\EndFor

	\State lastSeq \attr headPar, minValue \attr \maxLvl
	\State lock.release()
	\State \Return \textbf{false}
\EndIf

\State lastSeq \attr pred

\For{$i: \maxLvl \rightarrow 0$} \label{algMH:copyHead}
	\State headSeq[i] \attr headPar[i]
\EndFor

\State find(headSeq, lastSeq + 1, preds, succs) \label{algMH:unlink}

\For{$i: \maxLvl \rightarrow 0$}
	\State preds[i].next[i] \attr tail
	\State headPar.next[i] \attr succs[i]
\EndFor

\State lock.release()
\State \Return \textbf{true}
\end{algorithmic}
\end{algorithm}

Finally, the \texttt{SL::chopHead()} skiplist procedure is described in Alg.~\ref{Alg-ChopHead}. Note that all the \texttt{SL::find()} operations are executed outside the critical section. These operations identify the pointers that will change in order to relink the skiplist.

\begin{algorithm}[!htb]
\caption{SL::chopHead()}
\label{Alg-ChopHead}
\begin{algorithmic}[1]
\If{currSeq = \nullvalue}
	\State \Return \textbf{false}
\EndIf

\State find(headSeq, lastSeq.key + 1, preds, \nullvalue)
\State find(headSeq, currSeq.key, \nullvalue, succs)

\State lock.acquireW()

\For{$i: \maxLvl \rightarrow 0$}
	preds[i].next[i] \attr headPar.next[i]
\EndFor

\State lastSeq \attr headPar, currSeq \attr \nullvalue

\For{$i: \maxLvl \rightarrow 0$}
	\State headPar.next[i] \attr succs[i] \textbf{if} succs[i] $\ne$ tail
\EndFor

\State lock.release()

\State \Return \textbf{true}
\end{algorithmic}
\end{algorithm}

\section{Algorithms for Elimination and Combining}
\label{App-Algorithms-EliminationCombining}

In this section, we present the algorithms for the elimination and combining strategies for our priority queue, described in Section~\ref{Sec-Design-EliminationCombining}. The priority queue removal is shown in Alg.~\ref{Alg-removeMin}.


%

The priority queue insertion algorithm is shown in Alg.~\ref{Alg-add}. If the value being inserted is not suitable for the parallel part (\texttt{PQ::addPar()} returns false), the request is posted in the elimination array, until eliminated with a suitable \texttt{PQ::removeMin()} or consumed by the server thread. Details are discussed in Sec.~\ref{Sec-Design-EliminationCombining}. 

\begin{algorithm}[!htb]
\caption{PQ::add(inValue)}
\label{Alg-add}
\begin{algorithmic}[1]
\If{inValue $\le$ skiplist.minValue}
  \State rep \attr MAX\_ELIM\_MIN
\Else 
  \If{skiplist.addPar(inValue)}
    \State \Return $\mathbf{true}$
  \EndIf
  \State rep = MAX\_ELIM
\EndIf

\While{rep $> 0$}
	\State pos \attr $(id + 1) \% $ ELIM\_SIZE; (value, stamp) \attr elim[pos]
	\If {value = REMREQ \textbf{and} (inValue $\le$ skiplist.minValue))}
	    \If{CAS(elim[pos], (value, stamp), (inValue, 0))}
		\State \Return $\mathbf{true}$
	    \EndIf
	\EndIf
	\State rep \attr rep $- 1$; inc(pos)
\EndWhile

\If{skiplist.addPar(inValue)}
    \State \Return $\mathbf{true}$
\EndIf

\While{$\mathbf{true}$}
	\State (value, stamp) \attr elim[pos]
	\If {value = REMREQ \textbf{and} (inValue $\le$ skiplist.minValue))}
	    \If{CAS(elim[pos], (value, stamp), (inValue, 0))}
		\State \Return $\mathbf{true}$
	    \EndIf
	\EndIf
	\If {value = EMPTY}
	    \If{CAS(elim[pos], (value, stamp), (inValue, uniqueStamp()))}
		\Repeat
		  \State (value, stamp) \attr elim[pos]
		\Until{value = TAKEN}
		\State elim[pos] \attr (EMPTY, 0); \Return $\mathbf{true}$
	    \EndIf    
	\EndIf
	\State inc(pos)
\EndWhile
\end{algorithmic}
\end{algorithm}


\section{Implementing Combining and Elimination with Transactions}
\label{App-Transactions-Delegation}

In this section, we describe our experience using Intel TSX to simplify combining and elimination. Adapting the elimination algorithm to use transactions was straightforward, by just replacing the pessimistic synchronization with transactions. We note that a unique stamp as described in Section~\ref{Sec-Design-EliminationCombining} is not necessary for linearizability of elimination if the operations are performed inside hardware transactions. If a thread finds a matching operation and ensures in a transaction that the value is smaller than the minimum, then elimination is safe. If a change in the matching operation had occurred, the transaction would have aborted. We retry each transaction $N$ times (e.g. $N=3$ in our implementation). If a thread's transaction is aborted too many times during elimination, the thread moves on to other slots without retrying the failed slot in a fallback path. However, if the transaction fails while trying to insert an \texttt{PQ::add()} or \texttt{PQ::removeMin()} operation in an empty slot to be collected by the server thread, the original pessimistic algorithm is used as a software fallback path in order to guarantee forward progress. Unfortunately, the unique stamp needs to be used to ensure linearizability of the operations executed on the fallback path. 

Using transactions in the server thread implementation required including \texttt{SL::addSeq()} and \texttt{SL::removeSeq()} inside a transaction, which in turn caused too many aborts. Therefore, we designed an alternative combining algorithm that executes these operations outside the critical section. The complete algorithm is presented in Algorithm~\ref{Alg-Execute-OPT}. It is based on the observation that, as long as there is a sequential part in the skiplist, the \texttt{SL::removeSeq()} and the \texttt{SL::addSeq()} operations can be executed lazily. The server can use the skiplist's \emph{minValue} to return a value to a remove request and only execute the sequential operation after, without the remove thread waiting for it. Note that the skiplist's \texttt{minValue} could, in the meantime, return a value that is outdated. However, this value is always smaller or equal to the actual minimum in the skiplist, because it can only lag behind one sequential remove. This function is used by the \texttt{PQ::add()} operations to determine if they can eliminate or not. Therefore, estimating a minimum smaller than the actual minimum can affect performance, but will not impact correctness of our algorithm. Moreover, the server performs the \texttt{PQ::removeMin()} operation immediately after writing the minimum, thus cleaning up the sequential part and updating the minimum estimate. The \texttt{PQ::add()} case is similar too. If there is a sequential part to the skiplist, the server can update the skiplist lazily, after it releases the waiting thread. There is one difference. If the value inserted is smaller than \emph{minValue}, then this needs to be updated before releasing the waiting thread.

Using these changes in the combining algorithm allowed a straightforward implementation using hardware transactions. However, our experiments indicated that certain particularities of the best-effort HTM design make it unsuitable for this scenario. First of all, because of its  best-effort nature, a fallback is necessary in order to make progress. Therefore, the algorithm might be simplified on the common case, but it is still as complex as the fallback. Moreover, changes are often needed to adapt algorithms for an implementation using hardware transactions. Because these changes involve decreasing the sizes of the critical sections and decreasing the number of potential conflicts, these changes could be beneficial to the original algorithm too. Finally, it seems that communications paradigms, such as elimination and combining, are best implemented using pessimistic methods. Intel TSX has no means of implementing non-transactional operations inside transactions (also called escape actions) and no polite spinning mechanism to allow a thread to wait for a change that is going to be performed in a transaction. The spinning thread could often abort the thread that it is waiting for. We used the PAUSE instruction in the spinning thread to alleviate this issue, but better hardware support for implementing communication paradigms using hardware transactions is necessary. For our elimination and combining algorithms, we concluded that pessimistic synchronization works better. 

\begin{algorithm}[!htb]
\caption{Server::execute()}
\label{Alg-Execute-OPT}
\begin{algorithmic}[1]

\While{$\mathbf{true}$}
  \For{$i$: 1 $\rightarrow$ ELIM\_SIZE}
	\State (value, stamp) \attr elim[i]
	\If{value = REMREQ}
	    \If{skiplist.currSeq = \nullvalue}
		\State skiplist.moveHead()
	    \EndIf
	    \If{skiplist.currSeq $\ne$ \nullvalue}
		\If{CAS(elim[i], (value, stamp), (skiplist.minValue, 0))}
		  \State skiplist.removeSeq()
		\EndIf
	    \Else
	      \If{CAS(elim[i], (value, stamp), (INPROG, 0))}
		  \State min \attr skiplist.removeSeq(); elim[i] \attr (min, 0)
	      \EndIf
	    \EndIf
	\EndIf
	\If {IsValue(value) \textbf{and} (stamp $> 0$)}
	    \If{skiplist.currSeq $\ne$ \nullvalue}
		\If{CAS(elim[i], (value, stamp), (INPROG, 0))}
		  \If{value $<$ skiplist.minValue} \label{line:minValue}
		    \State skiplist.minValue \attr value
		  \EndIf
		  \State elim[i] \attr (TAKEN, 0); skiplist.addSeq(value)
		\EndIf
	    \Else
	      \If{CAS(elim[i], (value, stamp), (INPROG, 0))}
		\State skiplist.addSeq(value); elim[i] \attr (TAKEN, 0)
	      \EndIf
	    \EndIf

	\EndIf
  \EndFor
\EndWhile
\end{algorithmic}
\end{algorithm}

\end{document}